\documentclass{iau} 
\usepackage{graphicx}
\usepackage{amsmath}
\usepackage [english]{babel}
\usepackage [autostyle, english = american]{csquotes}
\MakeOuterQuote{"}

\title[UV P-Cygni Profiles of Magnetic Massive Stars] 
{New Insights into the Puzzling P-Cygni Profiles of Magnetic Massive Stars}

\author[Erba et al.]   
{Christiana Erba$^1$,  
	Alexandre David-Uraz$^{1,2}$, 
	V{\'e}ronique Petit$^{1,2}$,  
	\and Stanley P. Owocki$^1$  
}

\affiliation{
$^1$Deptartment of Physics and Astronomy, Bartol Research Institute, University of Delaware, Newark, DE, 19716, USA 
\\[\affilskip]
$^2$Department of Physics and Space Sciences, Florida Institute of Technology, Melbourne, FL 32904, USA
}

\pubyear{2017}
\volume{329}  
\jname{The Lives and Death-Throes of Massive Stars}
\editors{A.C. Editor, B.D. Editor \& C.E. Editor, eds.}

\begin{document}

\maketitle

\begin{abstract}
Magnetic massive stars comprise approximately 10\% of the total OB star population. Modern spectropolarimetry shows these stars host strong, stable, large-scale, often nearly dipolar surface magnetic fields of 1 kG or more. These global magnetic fields trap and deflect outflowing stellar wind material, forming an anisotropic magnetosphere that can be probed with wind-sensitive UV resonance lines. Recent HST UV spectra of NGC 1624-2, the most magnetic O star observed to date, show atypically unsaturated P-Cygni profiles in the C\textsc{iv} resonant doublet, as well as a distinct variation with rotational phase. We examine the effect of non-radial, magnetically-channeled wind outflow on P-Cygni line formation, using a Sobolev Exact Integration (SEI) approach for direct comparison with HST UV spectra of NGC 1624-2. We demonstrate that the addition of a magnetic field desaturates the absorption trough of the P-Cygni profiles, but further efforts are needed to fully account for the observed line profile variation. Our study thus provides a first step toward a broader understanding of how strong magnetic fields affect mass loss diagnostics from UV lines.
\end{abstract}

\firstsection

\section{Introduction}

Hot, luminous stars undergo mass loss through a steady outflow of supersonic material from the stellar surface. These radiatively driven \textit{stellar winds} are best characterized through modeling UV resonance lines, which are sensitive to wind properties. 

\begin{figure}[t!]
\centering
\includegraphics[scale=0.25]{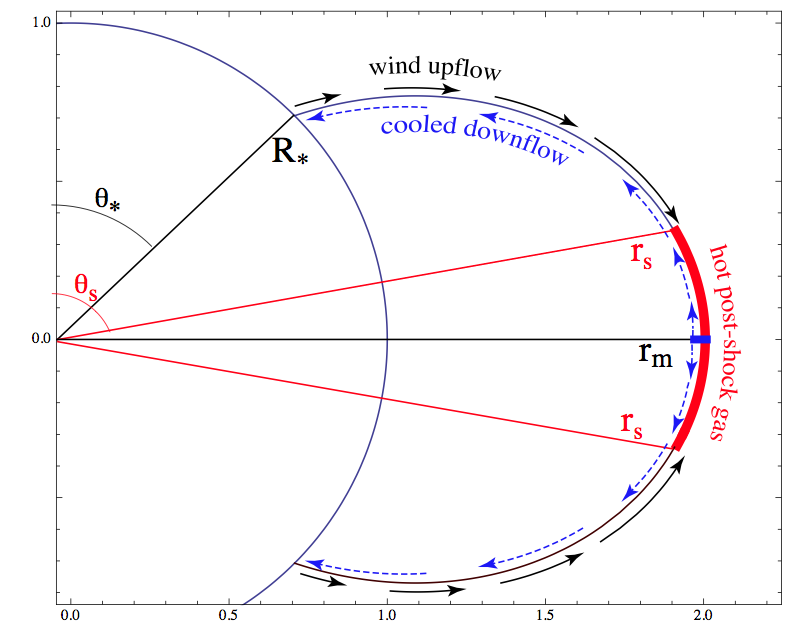}
\caption{Cartoon depiction of the Analytic Dynamical Magnetosphere (ADM) model, with a magnetic loop intersecting the stellar surface at colatitude $\theta_*$ and reaching a maximum radius $r_m$ at the magnetic equator. Magnetically-confined wind material from both stellar hemispheres flows up from the stellar surface (black arrows), and shocks at the magnetic equator. The shock front retreats to shock radius $r_s$, cooling the hot gas through the release of X-rays. The cooled post-shock material then flows back along the field lines toward the stellar surface (blue arrows). Reproduced from \cite[Owocki et al. (2016)]{owocki+2016}.}
\label{fig:adm_owocki}
\end{figure}

Recent spectropolarimetric measurements have revealed approximately 10\% of massive stars host strong, stable, nearly dipolar magnetic fields with surface field strength of approximately 1kG or more (\cite{wade+2016,fossati+2015}). The magnetic field channels the flow of the stellar wind along its field lines, confining the wind within closed loops and reducing the overall mass loss (see Fig. \ref{fig:adm_owocki}). This results in a \textit{dynamic} and \textit{structurally complex} so-called magnetosphere (\cite{uddoulaowocki2002}) with observational diagnostics in the optical, UV, and X-ray regimes (\cite{petit+2013}, David-Uraz et al., these proceedings). 

\begin{figure}[htb!]
\centering
\includegraphics[scale=0.3]{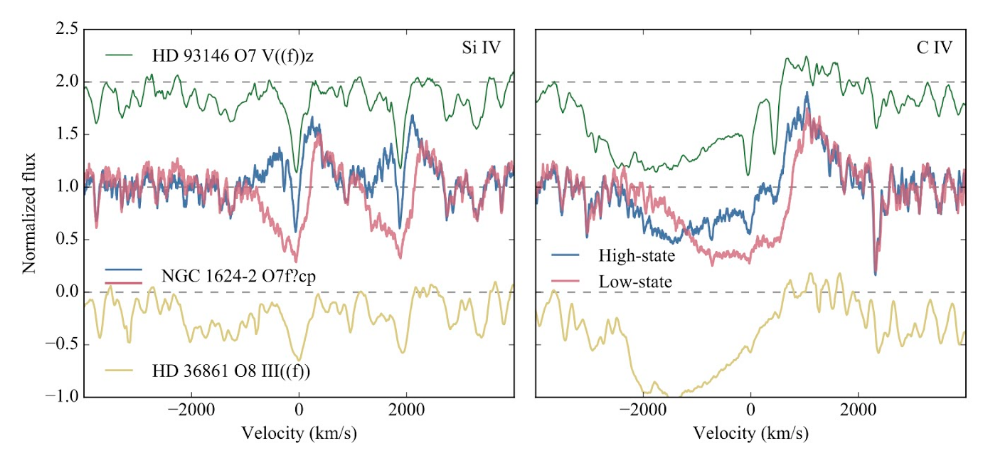}
\caption{HST/STIS data showing the comparison of the Si\textsc{iv} and C\textsc{iv} UV resonance lines of the magnetic O-star (NGC 1624-2, blue/red, middle) with non-magnetic O-stars of similar spectral type (HD 93146, green, top, and HD 36861, yellow, bottom). The absorption troughs of the C\textsc{iv} lines are a particularly illustrative example of the difference between non-magnetic O-stars (where the absorption trough is fully saturated) and magnetic O-stars (where the absorption trough is clearly unsaturated). Additionally, the magnetic pole-on view (blue) and equator-on view (red) of NCG 1624-2 are shown, demonstrating the modulation present in P-Cygni profiles of the same line at different phases.}
\label{fig:pcyg_real}
\end{figure}

Fig. \ref{fig:pcyg_real} shows HST UV spectra of NGC 1624-2, the most magnetic O star observed to date, compared with HD 93146 and HD 36861 (non-magnetic O stars of similar spectral type). In stark contrast to the strong line saturation observed in non-magnetic O stars, NGC 1624-2 and other magnetic O-type stars such as HD 57628, HD 191612, CPD -28 2561, show \textit{atypically unsaturated} P-Cygni profiles in the C\textsc{iv} resonant doublet, as well as a distinct dependence on rotational phase (\cite{grunhut+2009,wade+2011,wade+2012,marcolino+2013,naze+2015,marcolino+2012,petit+2013}). Therefore, the development of specialized analytical tools is necessary to interpret these profiles and derive the associated wind properties.

\section{The Magnetically Confined Wind}
 
In general, the dipolar axis of magnetic OB stars is not aligned with the rotational axis. Therefore, the resulting magnetospheres produce rotationally modulated variability in wind-sensitive UV line profiles and significantly reduced mass-loss rates (\cite{wade+2011b,sundqvist+2012,petit+2013}). P-Cygni line profiles can provide a powerful tool to probe the structure of these magnetospheres. A detailed investigation of the density and velocity structure of magnetically channeled winds has been carried out using magnetohydrodynamic (MHD) simulations (\cite{uddoulaowocki2002,uddoula+2008,uddoula+2009}). Although this work has provided unprecedented insight, these simulations remain computationally cumbersome and expensive.     

The recently published Analytic Dynamical Magnetosphere (ADM) model (\cite{owocki+2016}) provides a simplified parametric prescription corresponding to a time-averaged picture of the normally complex magnetospheric structure previously derived through full MHD simulations. As shown in Fig. \ref{fig:adm_owocki}, outflowing material leaves the stellar surface and is channeled along the field lines. At the magnetic equator, the outflowing material ("upflow") from each hemisphere collides, producing a shock. Cooling is achieved through X-ray production, allowing the wind material to flow back toward the stellar surface ("downflow"). This model shows remarkable agreement with H$\alpha$ and X-ray observations. However, as yet the ADM model has not been used to explain UV line variability, particularly with respect to the observed unsaturated absorption troughs of C\textsc{iv} lines. This provides the unique opportunity to conduct a thorough parameter study capable of accurately reproducing observed trends without the usual computational complexities associated with multi-dimensional MHD modeling.

\section{Initial Results and Future Applications}

We examine the effect of non-radial, magnetically channeled wind outflows on UV line formation through the development of synthetic UV wind-line profiles which implement the ADM formalism developed in \cite[Owocki et al. (2016)]{owocki+2016}. We solve the equation of radiative transfer using a Sobolev Exact Integration (SEI) method (both in the optically thin and optically thick approximations, following \cite[Owocki \& Rybicki (1985)]{owockirybicki1985}) to produce synthetic P-Cygni profiles for comparison with observational data. For simplicity, these preliminary models only implement the wind upflow component. We also limit these initial investigations to the magnetic "pole-on" and "equator-on" viewing angles.

\begin{figure}[htb!]
\centering
\includegraphics[scale=0.3]{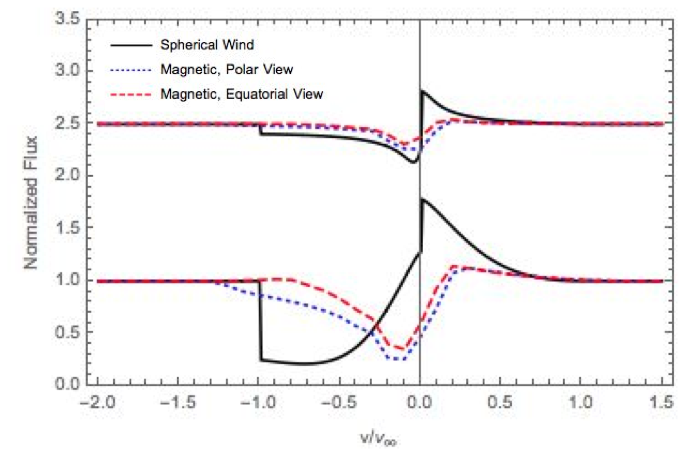}
\caption{Three synthetic wind-line profiles at line strengths similar to the C\textsc{iv} and Si\textsc{iv} UV resonance lines. The black curve shows the P-Cygni profile of a non-magnetic stellar wind with spherical symmetry; in contrast, the blue and red curves depict the P-Cygni profiles of a magnetically-confined wind viewed respectively from the magnetic pole-on and equator-on angles. These initial results suggest the addition of a dipole field alone is enough to produce unsaturated line profiles, although it does not reproduce the rotational modulation visible in HST/STIS observations.}
\label{fig:c_models}
\end{figure}

Initial results (see Fig. \ref{fig:c_models}) show that significant desaturation of the UV wind-line absorption troughs, both in the magnetic "pole-on" and "equator-on" views, occurs naturally from the addition of a dipolar magnetic field at the stellar surface. This is in agreement with earlier studies, which used MHD simulations to synthesize UV line profiles (\cite{marcolino+2013}) as well as observational data (see Fig. \ref{fig:pcyg_real}). This suggests that the desaturation is a direct result of the presence of a magnetic field. However, comparison with observational data also reveals a failure to reproduce the exact phase dependence of the lines, indicating the need for further investigation and additional modeling. 

In comparison to other computationally expensive MHD modeling techniques, the ADM formalism allows us to provide meaningful results that can be applied across a broader spectrum of magnetic massive stars. Future models will include both the wind upflow and downflow regions, as well as the addition of a “shock retreat” boundary to account for X-ray production in the wind. Further consideration of the effects of an approximated source function (as opposed to one that is solved self-consistently) also need to be explored. 

The detection and characterization of magnetic fields in massive stars currently relies heavily on spectropolarimetry, a powerful but expensive technique. Given these limitations, the future of the field of massive star magnetism might rely on indirect detection methods. Therefore, the development of robust UV diagnostics will be a critical step forward in understanding the circumstellar environment of massive stars, as well as the possible presence of an underlying field.



\begin{thebibliography}{}

\bibitem[Bard \& Townsend 2016]{bardtownsend2016}
{Bard, C. \& Townsend, R.H.D.} 2016, \textit{MNRAS}, 462.4, 3672

\bibitem[Fossati et al. 2015]{fossati+2015}
{Fossati, L., Castro, N., Sch{\"o}ller, M., Hubrig, S., Langer, N., Morel, T., Briquet, M., Herrero, A., Przybilla, N., Sana, H., et al.} 2015, \textit{A\&A}, 582, A45

\bibitem[Grunhut et al. 2009]{grunhut+2009}
{Grunhut, J.H., Wade, G.A., Marcolino, W.L.F., Petit, V., Henrichs, H.F., Cohen, D.H., Alecian, E., Bohlender, D., Bouret, J.C., Kochukhov, O., et al.} 2009, \textit{MNRAS}, 400.1, L94

\bibitem[Marcolino et al. 2012]{marcolino+2012}
{Marcolino, W.L.F., Bouret, J.C., Walborn, N.R., Howarth, I.D., Naz{\'e}, Y., Fullerton, A.W., Wade, G.A., Hillier, D.J., \& Herrero, A.} 2015, \textit{MNRAS}, 422.3, 2314

\bibitem[Marcolino et al. 2013]{marcolino+2013}
{Marcolino, W.L.F., Bouret, J.C., Sundqvist, J.O., Walborn, N.R., Fullerton, A.W., Howarth, I.D., Wade, G.A., \& ud-Doula, A.} 2013, \textit{MNRAS}, 431.3, 2253

\bibitem[Naz{\'e} et al. 2015]{naze+2015}
{Naz{\'e}, Y., Sundqvist, J.O., Fullerton, A.W., ud-Doula, A., Wade, G.A., Rauw, G., \& Walborn, N.R.} 2015, \textit{MNRAS}, 452.3, 2641

\bibitem[Owocki et al. 2016]{owocki+2016}
{Owocki, S.P., ud-Doula, A., Sundqvist, J.O., Petit, V., Cohen, D.H., \& Townsend, R.H.D.} 2016, \textit{MNRAS}, 462.4, 3830

\bibitem[Owocki \& Rybicki 1985]{owockirybicki1985}
{Owocki, S.P. \& Rybicki, G.B.} 1985, \textit{ApJ}, 299, 265

\bibitem[Petit et al. 2013]{petit+2013}
{Petit, V., Owocki, S.P., Wade, G.A., Cohen, D.H., Sundqvist, J.O., Gagn{\'e}, M., Apell{\'a}niz, J.M., Oksala, M.E., Bohlender, D.A., Rivinius, T., et al.} 2013, \textit{MNRAS}, 429.1, 398

\bibitem[Sundqvist et al. 2012]{sundqvist+2012}
{Sundqvist, J.O., ud-Doula, A., Owocki, S.P., Townsend, R.H.D., Howarth, I.D., \& Wade, G.A.} 2012, \textit{MNRAS}, 423.1, L21

\bibitem[Townsend \& Owocki 2005]{townsendowocki2005}
{Townsend, R.H.D. \& Owocki, S.P.} 2005, \textit{MNRAS}, 357.1, 251

\bibitem[ud-Doula et al. 2008]{uddoula+2008}
{ud-Doula, A., Owocki, S.P., \& Townsend, R.H.D.} 2008, \textit{MNRAS}, 385.1, 97

\bibitem[ud-Doula et al. 2009]{uddoula+2009}
{ud-Doula, A., Owocki, S.P., \& Townsend, R.H.D.} 2009, \textit{MNRAS}, 392.3, 1022

\bibitem[ud-Doula \& Owocki 2002]{uddoulaowocki2002}
{ud-Doula, A. \& Owocki, S.P.} 2002, 
\textit{ApJ}, 576.1, 413

\bibitem[Wade et al. 2011a]{wade+2011}
{Wade, G.A., Grunhut, J., Gr{\"a}fener, G., Howarth, I.D., Martins, F., Petit, V., Vink, J.S., Bagnulo, S., Folsom, C.P., Naz{\'e}, Y., et al.} 2011, \textit{MNRAS}, 419.3, 2459

\bibitem[Wade et al. 2011b]{wade+2011b}
{Wade, G.A., Howarth, I.D., Townsend, R.H.D., Grunhut, J.H., Shultz, M., Bouret, J.C., Fullerton, A., Marcolino, W., Martins, F., Naz{\'e}, Y., et al.} 2011, \textit{MNRAS}, 416.4, 3160

\bibitem[Wade et al. 2012]{wade+2012}
{Wade, G.A., Apell{\'a}niz, J.M., Martins, F., Petit, V., Grunhut, J., Walborn, N.R., Barb{\'a}, R.H., Gagn{\'e}, M., Garc{\'i}a-Melendo, E., Jose, J., et al.} 2012, \textit{MNRAS}, 425.2, 1278

\bibitem[Wade et al. 2016]{wade+2016}
{Wade, G. A., Neiner, C., Alecian, E., Grunhut, J.H., Petit, V., de Batz, B., Bohlender, D.A., Cohen, D.H., Henrichs, H.F., Kochukhov, O., et al.} 2016, \textit{MNRAS}, 456.1, 2

\end{thebibliography}

\end{document}